\documentstyle[12pt]{article}

\catcode`\@=11
\def\@citex[#1]#2{%
\if@filesw \immediate \write \@auxout {\string \citation {#2}}\fi
\@tempcntb\m@ne \let\@h@ld\relax \def\@citea{}%
\@cite{%
  \@for \@citeb:=#2\do {%
    \@ifundefined {b@\@citeb}%
      {\@h@ld\@citea\@tempcntb\m@ne{\bf ?}%
      \@warning {Citation `\@citeb ' on page \thepage \space undefined}}%
      {\@tempcnta\@tempcntb \advance\@tempcnta\@ne%
      \@tempcntb\number\csname b@\@citeb \endcsname \relax%
      \ifnum\@tempcnta=\@tempcntb 
	\ifx\@h@ld\relax%
	  \edef \@h@ld{\@citea\csname b@\@citeb\endcsname}%
	\else%
	  \edef\@h@ld{\ifmmode{-}\else--\fi\csname b@\@citeb\endcsname}%
	\fi%
      \else
	\@h@ld\@citea\csname b@\@citeb \endcsname%
	\let\@h@ld\relax%
      \fi}%
    \def\@citea{,\penalty\@highpenalty\,}%
  }\@h@ld
}{#1}}

\def\@citeb#1#2{{[#1]\if@tempswa , #2\fi}}
\def\@citeu#1#2{{$^{#1}$\if@tempswa , #2\fi }}
\def\@citep#1#2{{#1\if@tempswa , #2\fi}}

\def\bcites{         
	\catcode`\@=11
	\let\@cite=\@citeb
	\catcode`\@=12
}

\def\upcites{         
	\catcode`\@=11
	\let\@cite=\@citeu
	\catcode`\@=12
}

\def\plaincites{      
	\catcode`\@=11
	\let\@cite=\@citep
	\catcode`\@=12
}

\newcount\hour
\newcount\minute
\newtoks\amorpm
\hour=\time\divide\hour by 60
\minute=\time{\multiply\hour by 60 \global\advance\minute by-\hour}
\edef\standardtime{{\ifnum\hour<12 \global\amorpm={am}%
	\else\global\amorpm={pm}\advance\hour by-12 \fi
	\ifnum\hour=0 \hour=12 \fi
	\number\hour:\ifnum\minute<10 0\fi\number\minute\the\amorpm}}
\edef\militarytime{\number\hour:\ifnum\minute<10 0\fi\number\minute}

\def\draftlabel#1{{\@bsphack\if@filesw {\let\thepage\relax
   \xdef\@gtempa{\write\@auxout{\string
      \newlabel{#1}{{\@currentlabel}{\thepage}}}}}\@gtempa
   \if@nobreak \ifvmode\nobreak\fi\fi\fi\@esphack}
	\gdef\@eqnlabel{#1}}
\def\@eqnlabel{}
\def\@vacuum{}
\def\marginnote#1{}
\def\draftmarginnote#1{\marginpar{\raggedright\scriptsize\tt#1}}
\def\draft{
	\pagestyle{plain}
	\overfullrule=2pt
	\oddsidemargin -.5truein
	\def\@oddhead{\sl \phantom{\today\quad\militarytime} \hfil
	\smash{\Large\sl DRAFT} \hfil \today\quad\militarytime}
	\let\@evenhead\@oddhead
	\let\label=\draftlabel
	\let\marginnote=\draftmarginnote
	\def\ps@empty{\let\@mkboth\@gobbletwo
	\def\@oddfoot{\hfil \smash{\Large\sl DRAFT} \hfil}
	\let\@evenfoot\@oddhead}
	\def\@eqnnum{(\theequation)\rlap{\kern\marginparsep\tt\@eqnlabel}%
	\global\let\@eqnlabel\@vacuum}  }

\def\blackfonts{
	\font\blackboard=msbm10 scaled\magstep1
	\font\blackboards=msbm8
	\font\blackboardss=msbm6
}

\def\nblack{            
	\def\ZZ{{Z \n{10} Z}}
	\def\NN{{N \n{14} N}}
	\def\CC{{C \n{11} C}}
	\def\RR{{R \n{11} R}}
	\def\QQ{{Q \n{12} Q}}
	\def\PP{{P \n{11} P}}
}

\def\prep{         
	\catcode`\@=11
	\input art10.sty
	\catcode`\@=12
	
	\let\small\null
	\def\blackfonts{
		\font\blackboard=msbm10
		\font\blackboards=msbm7
		\font\blackboardss=msbm5
	}
	\let\sl\it
	\twocolumn
	\sloppy
	\voffset=-2.54truecm
	\hoffset=-2.54truecm
	\flushbottom
	\parindent 1em
	\leftmargini 2em
	\leftmarginv .5em
	\leftmarginvi .5em
	\marginparwidth 48pt
	\marginparsep 10pt
	\setlength{\columnsep}{2truecm}
	\setlength{\textwidth}{25.4truecm}
	\setlength{\textheight}{17truecm}
	\baselineskip=16pt
	\oddsidemargin .18truein
	\evensidemargin .17truein
}

\def\eqalign#1{\null\,\vcenter{\openup\jot\m@th
  \ialign{\strut\hfil$\displaystyle{##}$&$\displaystyle{{}##}$\hfil
      \crcr#1\crcr}}\,}
\def\eqalignno#1{\displ@y \tabskip\centering
  \halign to\displaywidth{\hfil$\@lign\displaystyle{##}$\tabskip\z@skip
    &$\@lign\displaystyle{{}##}$\hfil\tabskip\centering
    &\llap{$\@lign##$}\tabskip\z@skip\crcr
    #1\crcr}}

\def\section{\@startsection {section}{1}{\z@}{3.ex plus 1ex minus
 .2ex}{2.ex plus .2ex}{\large\bf}}
\def\subsection{\@startsection{subsection}{2}{\z@}{2.75ex plus 1ex minus
 .2ex}{1.5ex plus .2ex}{\bf}}

\def\appendix{{\newpage\section*{Appendices}}\let\appendix\section%
	{\setcounter{section}{0}
	\gdef\thesection{\Alph{section}}}\section}

\def\abstract{\if@twocolumn
\section*{Abstract}
\else 
\begin{center}
{\bf Abstract\vspace{-.5em}\vspace{0pt}}
\end{center}
\quotation
\fi}

\catcode`\@=12

\def\noj#1,#2,{{\bf #1} (19#2)\ }
\def\jou#1,#2,#3,{{\sl #1\/ }{\bf #2} (19#3)\ }
\def\ann#1,#2,{{\sl Ann.\ Physics\/ }{\bf #1} (19#2)\ }
\def\cmp#1,#2,{{\sl Comm.\ Math.\ Phys.\/ }{\bf #1} (19#2)\ }
\def\cq#1,#2,{{\sl Class.\ Quantum Grav.\/ }{\bf #1} (19#2)\ }
\def\cqg#1,#2,{{\sl Class.\ Quantum Grav.\/ }{\bf #1} (19#2)\ }
\def\ijmp#1,#2,{{\sl Int.\ J.\ Mod.\ Phys.\/ }{\bf A#1} (19#2)\ }
\def\jmp#1,#2,{{\sl J.\ Math.\ Phys.\/ }{\bf #1} (19#2)\ }
\def\grg#1,#2,{{\sl Gen.\ Rel.\ Grav.\/ }{\bf #1} (19#2)\ }
\def\mpl#1,#2,{{\sl Mod.\ Phys.\ Lett.\/ }{\bf A#1} (19#2)\ }
\def\nc#1,#2,{{\sl Nuovo Cim.\/ }{\bf #1} (19#2)\ }
\def\np#1,#2,{{\sl Nucl.\ Phys.\/ }{\bf B#1} (19#2)\ }
\def\pl#1,#2,{{\sl Phys.\ Lett.\/ }{\bf #1B} (19#2)\ }
\def\pla#1,#2,{{\sl Phys.\ Lett.\/ }{\bf #1A} (19#2)\ }
\def\pr#1,#2,{{\sl Phys.\ Rev.\/ }{\bf #1} (19#2)\ }
\def\prd#1,#2,{{\sl Phys.\ Rev.\/ }{\bf D#1} (19#2)\ }
\def\prl#1,#2,{{\sl Phys.\ Rev.\ Lett.\/ }{\bf #1} (19#2)\ }
\def\prp#1,#2,{{\sl Phys.\ Rept.\/ }{\bf #1C} (19#2)\ }
\def\ptp#1,#2,{{\sl Prog.\ Theor.\ Phys.\/ }{\bf #1} (19#2)\ }
\def\ptpsup#1,#2,{{\sl Prog.\ Theor.\ Phys.\/ Suppl.\/ }{\bf #1} (19#2)\ }
\def\rmp#1,#2,{{\sl Rev.\ Mod.\ Phys.\/ }{\bf #1} (19#2)\ }
\def\yadfiz#1,#2,#3[#4,#5]{{\sl Yad.\ Fiz.\/ }{\bf #1} (19#2) #3%
\ [{\sl Sov.\ J.\ Nucl.\ Phys.\/ }{\bf #4} (19#2) #5]}
\def\zh#1,#2,#3[#4,#5]{{\sl Zh.\ Exp.\ Theor.\ Fiz.\/ }{\bf #1} (19#2) #3%
\ [{\sl Sov.\ Phys.\ JETP\/ }{\bf #4} (19#2) #5]}

\def\beq{\begin{equation}}
\def\eeq{\end{equation}}
\def\beqar{\begin{eqnarray}}
\def\eeqar{\end{eqnarray}}

\def\nfrac#1#2{{\displaystyle{\vphantom1\smash{\lower.5ex\hbox{\small$#1$}}%
	\over\vphantom1\smash{\raise.25ex\hbox{\small$#2$}}}}}

\def\p#1{\mskip#1mu}
\def\n#1{\mskip-#1mu}
\def\stop{\p6.}
\def\comma{\p6,}

\def\lae{\mathrel{\mathop{\smash{\lower .5 ex \hbox{$\stackrel<\sim$}}}}}
\def\lae{\mathrel{\mathop{\smash{\lower .5 ex \hbox{$\stackrel>\sim$}}}}}

\def\l:{\mathopen{:}\,}
\def\r:{\,\mathclose{:}}

\def\[{\left[}          \def\]{\right]}
\def\({\left(}          \def\){\right)}
\def\<{\left<}          \def\>{\right>}

\def\CF{{\cal F}}
\def\CL{{\cal L}}

\catcode`\@=11
\def\theequation{\arabic{equation}}
\catcode`\@=12

\nblack
\bcites

\nblack

\catcode`\@=11
\def\theequation{\thesection.\arabic{equation}}
\@addtoreset{equation}{section}
\@addtoreset{footnote}{section}
\@addtoreset{footnote}{subsection}
\catcode`\@=12

\newcommand{\beqn}{\begin{equation}}
\newcommand{\eeqn}{\end{equation}}
\newcommand{\beqnarray}{\begin{eqnarray}}
\newcommand{\eeqnarray}{\end{eqnarray}}

\newcommand{\res}{\;\mathop{\mbox{\rm res}}}
\begin{document}
\begin{titlepage}

\begin{center}
Oct 2, 1995
\hfill       IASSNS-HEP-95/76 \\
\hfill                  hep-th/9509176

\vskip 1 cm
{\large \bf On the
Quantum Moduli Space of Vacua of $N=2$ Supersymmetric Gauge Theories\\}
\vskip 0.1 cm
\vskip 0.5 cm
{Amihay Hanany
}
\vskip 0.2cm
{\sl
hanany@sns.ias.edu \\
School of Natural Sciences \\
Institute for Advanced Study \\
Olden Lane, Princeton, NJ 08540, USA
}

\end{center}

\vskip 0.5 cm
\begin{abstract}
Families of hyper-elliptic curves which describe the quantum
moduli spaces of vacua of $N=2$ supersymmetric $SO(N_c)$ gauge theories
coupled to $N_f$ flavors of quarks in the vector representation are
constructed. The quantum moduli spaces for $2N_f < N_c-1$ are
determined completely by imposing $R$-symmetry,
instanton corrections and
the proper classical singularity structure. These curves are verified
by residue calculations.
The quantum moduli spaces
for $2N_f\geq N_c-1$ theories are parameterized and their general
structure is worked out using residue calculations.
The exact metrics on the quantum moduli spaces as well as the exact
spectrum of stable massive states are derived.
The results presented here together with recent results of Martinec
and Warner provide a natural conjecture for the form of the curves for
the other gauge groups.
\end{abstract}

\end{titlepage}

\section{Introduction}

There has been much progress recently in the study of
non-perturbative properties of $N=2$
four dimensional supersymmetric gauge theories.
Seiberg and Witten \cite{sw1} found an exact soultion of the low
energy effective
action as well as discovered a reach structure of duality phenomena.
This work was generalized in \cite{sw2} to include massless and massive
matter hyper-multiplets.
The quantum moduli space of $N=2$ supersymmetric pure $SU(N_c)$
gauge theories was constructed as an $N_c-1$
parameter family of hyper-elliptic curves in \cite{Shimon,Alon},
and the physics of the model was studied in \cite{DS,AD}.
The inclusion of matter multiplets was carried out in \cite{HO,APS}
and was also considered in \cite{MN}.
The curves for $SO(2l+1)$ gauge groups were constructed in \cite{Dan}
and for $SO(2l)$ gauge groups in \cite{BL}.

Our aim in this paper is to determine the quantum moduli spaces of
vacua of the Coulomb phase of $N=2$ supersymmetric $SO(N_c)$ gauge
theories with $N_f$ matter
hyper-multiplets in the vector representation, as well as
the exact metric on the quantum moduli space and the exact spectrum of
massive stable particles.
The quantum moduli spaces will be constructed as families of
hyper-elliptic
curves satisfying a set of physical constraints: $R$-symmetry,
singularity structure and global symmetries,
appropriate inclusion of instanton correction, proper classical
and scaling (integration of massive quarks) limits,
compatibility of residue calculations with the BPS formula and
correct weak coupling monodromies.

The paper is organized as follows. In section two
we review general aspects of $N=2$ gauge
theories and the moduli spaces of vacua,
introduce the physical quantities
associated with them,
discuss their singularity
structure and present principles for their construction.
In section three we construct the quantum moduli space for
$N_f<N_c-l-2$.
We show that it is determined completely by imposing $R$-symmetry,
instanton corrections and
the proper classical singularity structure.
We get compatibility of the residue calculations with the BPS formula
which provides a consistency check of the result.
The begining of section four is devoted to the $N_f=N_c-l-2$ theories.
Applying the physical constraints including residue calculations
leaves us with one undetermined constant parameter.

In the rest of section four the quantum moduli spaces
for $N_f > N_c-l-2$ theories are parameterized and their general
structure is worked out using
residue calculations. This still leaves us with undetermined
constant coefficients.
Global symmetry considerations may suggest a complete determination
of them. We absorb the factors into redefinition of the symmetric
polynomials.
Weak coupling
monodromy calculations along the lines of \cite{HO} give support to the
results in the previous sections.

In section five we discuss the $N_f=N_c-2$ theories.
When the bare masses and all but one of the moduli are zero
one expects scale invariant theories
with periods satisfying the classical relations.
The quantum moduli spaces in the massless as well as
the massive theories are parameterized and their general
structure is worked out using
residue calculations in a similar manner to the
$N_f>N_c-l-2$ case.
Again, global symmetry considerations may suggest a complete description
but we do not find such yet. We absorb them into redefinition of
the moduli.

In general our results provide the exact metrics on the
quantum moduli spaces as well as the exact
spectrum of stable massive states.
Section six is devoted to discussion and conclusions.
We outline there a conjecture for the curves to any simple group.

\section{The moduli space of vacua of $N=2$ gauge theories}
In this section we review general aspects of $N=2$ gauge
theories and the moduli spaces of vacua,
introduce the physical quantities
associated with them,
discuss their singularity
and monodromy structures,
and present principles for their construction.

\subsection{$N=2$ QCD and the moduli space of vacua}

We will consider $N=2$ supersymmetric gauge theories
with gauge group $G=SO(N_c)$ and $N_f$ flavors.
The field content of the theories consists of $N=2$ chiral multiplet
and $N_f$ hyper-multiplets.
The $N=2$ chiral multiplet contains gauge fields $A_{\mu}$, two
Weyl fermions $\lambda$ and $\psi$ (alternatively one Dirac fermion) and a
complex scalar $\phi$, all in the
adjoint representation of the gauge group G.
The $N=2$ hyper-multiplets that we will consider
contain two Weyl fermions $\psi_q$ and $\psi_{\tilde{q}}^{\dagger}$ and two
complex bosons $q$ and $\tilde{q}^{\dagger}$, all in the vector
representation $r$ of $G$.

In terms
of $N=1$ super-fields the $N=2$
chiral multiplet consists of a vector multiplet $W_{\alpha}$ and a
chiral multiplet  $\Phi$, while the $N=2$
hyper-multiplets consist of two chiral super-fields
$Q^i{}_a$ and $\tilde Q_i{}_a$, where $i=1,\dots,N_f$ is
the flavor index and
$a=1,\dots,N_c$ is the color index.
The super-potential in the $N=1$ language reads
\beq
W=\sqrt {2} \tilde Q_i \Phi Q^i + \sum_i m_i\tilde Q_i Q^i  \comma
\label{superpot}
\eeq
with $m_i$ being the bare masses and color indices are suppressed.
The first term in (\ref{superpot}) is related  by $N=2$
super-symmetry to the gauge couplings and the second term corresponds
to $N=2$ invariant mass terms.

The theory has an $l$ complex dimensional moduli space of vacua,
which are parameterized by the gauge invariant order parameters
\beq
u_{2k} = Tr\langle\phi^{2k}\rangle,~~~~~~ k=1,\ldots,l
\comma
\label{invar}
\eeq
$\phi$ being the
scalar field of the $N=2$ chiral multiplet.
$l={N_c\over2}$ for $N_c$ even and $l={N_c-1\over2}$ for $N_c$ odd.
The moduli space of vacua corresponds to $\phi$
satisfying a D-flatness condition $[\phi,\phi^\dagger]=0$.
Thus, up
to a gauge transformation we can take $\phi$ to be a block diagonal
matrix,
\beq
\langle \phi \rangle =
diag[A_1,\ldots,A_{l}] \comma
\label{vev}
\eeq
where $A_i=\pmatrix{0 & a_i \cr -a_i & 0 \cr}$.
At weak coupling we have
\beq
u_{2k} = \sum_{i=1}^{l} a_i^{2k} \stop
\label{uka}
\eeq

Alternative gauge invariant order parameters
are defined at the classical level as the
symmetric polynomials $s_{2k}$ in $a_i^2$
\beq
s_{2k} = (-)^k \sum_{i_1<\cdots<i_k} a_{i_1}^2\cdots a_{i_k}^2,
{}~~~~~ k=1,\ldots,l
\comma
\label{symm}
\eeq

Using (\ref{uka}) and (\ref{symm}) the sets $s_k$ and $u_k$ are
related by Newton's formula
\beq
ks_k + \sum_{i=1}^k s_{k-i}u_i = 0 ,~~~~~~ k=\{1,2,\ldots\} \comma
\label{Newton}
\eeq
where $s_0=1,s_{2k+1}=u_{2k+1}=0$.
This relation serves as a definition of $s_k$ at the quantum level.

At a generic point the expectation values of $\phi$ break
the gauge symmetry to $U(1)^l$ and a low-energy effective Lagrangian can
be written in terms of multiplets $(A_i,W_i)$, $i=1,\ldots,l$.

The $N=2$ effective Lagrangian
takes the form
\beq
\CL_{eff}=Im{\frac{1}{4\pi}}\[\int d^4\theta\partial_i\CF(A)\bar
 A^i+{1\over2}\int
d^2\theta\partial_i\partial_j\CF(A)W^iW^j\] \comma
\label{oneloop}
\eeq
where $\CF$ is a holomorphic pre-potential.

$N=2$ super-symmetry
implies that there are no perturbative corrections beyond one loop.
There are, however, non-perturbative instanton corrections.

The classical moduli space is described by the family of genus $g=2l-1$
hyper-elliptic curves given by \cite{Dan,BL}
\beq
y^2={\cal C}^2(x),\qquad {\cal C}(x) =
\det(x I_c-\phi)  \comma
\label{cl1}
\eeq
where $I_c$ is the $N_c\times N_c$ identity matrix
and in terms of the symmetric functions $s_i$
\beq
{\cal C}(x)=x^{2l}+\sum_{i=1}^ls_{2i} x^{2(l-i)}
\label{cl}
\eeq
The singularities of the classical curve correspond to cases where the gauge
group is spontaneously broken to non-Abelian subgroups of $G$ rather than
to $U(1)^{l}$.

We will construct the quantum moduli spaces of vacua
for the $SO(N_c)$ groups (For other
groups we will use a description recently advocated by Martinec and
Warner \cite{MW}.) as families of hyper-elliptic curves
\beq
y^2 = \prod_{i=1}^{4l}(x-e_i) \comma
\label{hyper}
\eeq
where the roots $e_i$ are functions of the quark masses $m_i$,
the dynamically generated
scale $\Lambda$
\footnote{When $2h=2N_f\mu$, the role of the dynamical scale $\Lambda$
is played by a modular form of the appropriate modular group. $h$ is the
second Casimir and $\mu$ is the Dynkin index of the representation. For
$SO(N_c)$ $h=N_c-2$, $\mu=1$}
and the
gauge invariant order parameters
$u_k$ or $s_k$, with (\ref{cl1}) being their classical limit.

\subsection{Dyon spectrum and duality}
The spectrum of $N=2$ supersymmetric QCD includes
particles in the ``small" representation of the $N=2$ algebra,
the so called
BPS-saturated states. These are electrically
and magnetically charged particles
with masses
\beq
M^2=2|Z|^2 \comma
\label{MZ}
\eeq
where $Z$ is a central extension of the $N=2$ supersymmetry
algebra and is a linear combination of conserved charges \cite{wo}.
It reads \cite{sw2}
\beq
Z = \sum_{i=1}^{l}[n_e^i a^i +n_m^i a_D^i] +\sum_{i=1}^{N_f}S_i m_i \comma
\label{BPS}
\eeq
where $n_e,n_m$ are the electric and magnetic charges.
$S_i$ are the $U(1)$ charges corresponding to additional symmetries
that may exist  when the global symmetry is
explicitly broken by non-zero masses.
$m_i$ are the bare masses of the hyper-multiplets
and  $a,a_D$ correspond to the vacuum expectation values of the
scalar component of the chiral super-field $A$ and its dual
\beq
a_D^i=\frac{\partial\CF(a)}{\partial a} \stop
\label{ad}
\eeq

Mathematically,
$a,a_D$ are sections of a certain bundle over the moduli space
of vacua.
When all the masses $m_i$ are zero, $a,a_D$ are sections
of an $Sp(2l,Z)$ bundle, thus their monodromies upon traversing a
closed cycle in the moduli space are elements of
$Sp(2l,Z)$.
The mass formula (\ref{MZ}) is
$Sp(2l,Z)$ invariant, reflecting a duality property of the theory.
When the masses do not vanish, constant shifts are
allowed transformations in addition to the above $Sp(2l,Z)$
structure \cite{sw1,sw2}. Thus, the monodromy transformation
takes the form
\beq
\pmatrix{a_D \cr a}\rightarrow M\pmatrix{a_D \cr a}+c,~~~~~~~~M
\in Sp(2l,Z)\comma
\label{mondef}
\eeq
where $c$ is independent of the moduli parameters
and depends on the masses of the quarks.

$a,a_D$ can be written as periods of a meromorphic one form $\lambda$
on the curve describing the space of vacua
\beq
a_D^i=\oint_{\alpha_i}\lambda \comma~~~~~~
a^i=\oint_{\beta_i}\lambda \comma
\label{aad}
\eeq
where $\alpha_i,\beta_j$ form a basis of homology cycles on the curve.
In order to determine $\lambda$ one assumes that
\beqar
\frac{\partial a_D^i}{\partial s_{2k}}
& \propto &\oint_{\alpha_i}x^{2(l-k-1)}\frac{d x}{y}
 ~~~~~~k=1,\ldots,l\nonumber\\
\frac{\partial a^i}{\partial s_{2k}}
& \propto &\oint_{\beta_i}x^{2(l-k-1)}\frac{d x}{y}~~~~~~k=1,\ldots,l \comma
\label{aap}
\eeqar
with  $x^{2(l-k-1)}\frac{d x}{y}~~~k=1,\ldots,l$
being a basis of holomorphic one forms on the curve.
The exact proportionality factor in (\ref{aap}) is determined
by matching $a^i$ to the weak coupling relations (\ref{uka})
or (\ref{symm}).
It is a constant since we want to avoid unwanted zeros or
poles.
Its value
is ${1\over2\pi i}$.

Note that (\ref{aad}) and (\ref{aap}) yield, up to an exact form,
\beq
\frac{d \lambda}{d s_k} \propto x^{2(l-k-1)}\frac{d
x}{y}~~~~~~k=1,\ldots,l\stop
\label{lambda}
\eeq

When the bare masses are zero the residues of $\lambda$
vanish, thus ensuring that
 $a,a_D$ are invariant under deformations of the cycle of integration
even across poles of $\lambda$.
For non zero masses the residues take the form
\beq
2 \pi i \res(\lambda) = \sum_in_im_i ~~~~~~ n_i \in Z \comma
\label{res}
\eeq
thus allowing jumps in $a,a_D$, as defined in (\ref{aad}),
when crossing poles of $\lambda$ compatible with the mass formula
(\ref{BPS}).

The matrix of low energy coupling constants, $\tau$, is given by
\beq
\tau_{ij}(a)= \frac{\partial a_D^i}{\partial a^j} \stop
\label{pmatrix}
\eeq
By virtue of (\ref{aap}), $\tau_{ij}$ is the period matrix of the curve
describing the quantum moduli space.
The metric on quantum moduli space reads
\beq
(d s)^2 = Im~ d a_D^i d\bar{a}^i
\comma
\label{metric}
\eeq
is invariant under the transformation (\ref{mondef}),
and is positive definite.

\subsection{The singularity structure}

As discussed in section 2.1, the $l$ dimensional
moduli space of vacua ${\cal M}_{l}$ is parameterized
by gauge invariant order parameters such as
$s_k$ of equation (\ref{symm}).
The singular locus of the family of curves (\ref{hyper})
describing the moduli space
is the codimension one variety defined as the
vanishing locus of the discriminant
\footnote{When the coefficient of the highest order monomial $x^n$
in the polynomial is
$a \neq 1$ the discriminant (\ref{var}) is modified by a pre-factor
$a^{2n-2}$.}
\beq
\Delta[s_k] \equiv \prod_{i<j}(e_i-e_j)^2 = 0 \stop
\label{var}
\eeq
Along this variety additional massless states appear in the spectrum and the
effective low energy description (\ref{oneloop})
is not valid.

In the semi-classical limit $\Lambda\rightarrow 0$ the discriminant
factorizes
\beq
\Delta[\Lambda\rightarrow 0] = \Lambda^{h_g(2h-2N_f\mu)}
\Delta_{c}^2\Delta_{f} \comma
\label{Del}
\eeq
with $h_g=2l$ for $SO(N_c)$. where
\beq
\Delta_{c} = \Delta({\cal C}) \comma
\label{D0}
\eeq
is the discriminant of ${cal C}(x)$ and
\beq
\Delta_f=\prod_{j=1}^{N_f}{\cal C}^{2\mu}(x^2=m_i^2).
\label{D1}
\eeq
The zero locus of $\Delta_{c}$ corresponds to singularities
where classically the group $G$ is not spontaneously broken
to $U(1)^l$ but rather to a larger subgroup (In particular,
the gauge group is not broken at all at the origin of the moduli space).
The zero locus of $\Delta_{f}$ defines a complex codimension one
variety in the moduli space where a quark becomes massless
classically.
The product up to $N_f$ in (\ref{D1}) is a consequence of having $N_f$
different flavors.
The power of $\Lambda$ in (\ref{Del}) follows from instanton contribution and
from the fact that classically $N_c$ roots degenerate.

\subsection{Principles for the construction of the moduli spaces}

In the following we summarize the
principles which will be used by us in the construction of the
families of hyper-elliptic curves describing the quantum
moduli spaces of vacua:

\begin{enumerate}
\item {{\bf Symmetry} : The curves are invariant under $R$ charge
transformation
\beq
O \rightarrow \exp\[\frac{2 \pi i R(O)}{4(2h-2N_f\mu)
)}\] O \comma
\label{Rs}
\eeq
where $R(O)$ is the $R$ charge of $O$ and $O$ refers to the
building blocks of the curves, $y,x,m_i,\Lambda,s_k$.
We have
\begin{center}
\label{Rc}
\begin{tabular}{|c||c|c|c|c|c|c|}
\hline
$O$ & $y$ & $x$ & $m_i$ & $\Lambda$ & $u_k$ & $s_k$ \\ \hline
$R(O)$ & $4l$ & $2$ & $2$ & $2$ & $2k$ & $2k$ \\ \hline
\end{tabular}
\end{center}
}

\item {{\bf Singularity structure} : As discussed in section
2.3, singularities of a curve
describing a quantum moduli space
are associated with the zero locus of the corresponding
discriminant and are physically interpreted as
the appearance of additional massless states in the spectrum,
where the order of vanishing of the discriminant at a point corresponding
to a singularity of the curve is generically the number of
massless states at that point.
These states  belong to a representation of the global symmetry
group and the order of vanishing of the discriminant is
its dimension. Thus, global symmetry imposes constraints on the
singularity structure.}

\item {{\bf Instanton corrections} : The one instanton process contribution
to the curves for $2N_f\mu <2h$ takes the form \cite{nati}
\beq
\Lambda^{2h-2N_f\mu} \stop
\label{instant}
\eeq
}

\item {{\bf Integration of a massive quark}:
Sending a quark mass $m\rightarrow \infty$ in $N_f\mu < h$
theories,
and sending the scale $\Lambda_{N_f} \rightarrow 0$ such that
\footnote{In \cite{FP} this renormalization scheme was identified as the
$\overline{DR}$ scheme.}
\beq
\Lambda_{N_f-1}^{2h-2(N_f-1)\mu} = m^{2\mu}\Lambda_{N_f}^{2h-2N_f\mu}
\comma
\eeq
is fixed,
we reduce the number of flavors from
$N_f$ to $N_f-1$ and we require compatibility of the curves.
When $2N_f\mu=2h$ the appropriate matching condition reads
\beq
\Lambda_{2h-1}^{2\mu} = m^{2\mu} q \comma
\label{ql}
\eeq}
where $q \equiv e^{2\pi i\tau}$.
\item {{\bf Classical limit}: The curves should exhibit the singularity
structure (\ref{Del})-(\ref{D1})
at the classical level
$\Lambda \rightarrow 0$.
}
\item {{\bf Residues} :
As discussed in section 2.2,
the meromorphic one form on the curve $\lambda$ in (\ref{aad})
may have poles but its residues are restricted
by (\ref{res}).
This, as we shall see, leads to powerful constraints on the
structure of the curves.
}
\end{enumerate}

Our aim is to construct the hyper-elliptic curves describing the
quantum moduli spaces of $N=2$ QCD with $N_c$ colors and $N_f$
flavors in the vector representation
of the gauge group. For this, the form of the hyper-elliptic curves introduced
in \cite{Dan,BL} is the appropriate framework.
The general strategy for constructing the curves will be to first
restrict the
possible terms to those compatible with $R$ charge symmetry and the
form of the instanton corrections. We then impose the proper
classical singularity structure and use residue calculations.

Since the one loop beta function of the theory is proportional to
$2h-2N_f\mu$
(higher loop corrections to it vanish) we limit ourselves to
$N_f\mu<h$ where the
theory is asymptotically free and to $N_f\mu=h$ where the
(massless) theory is scale
invariant.

\section{The quantum moduli space for  $N_f\mu <h-l$}

\subsection{The general case}
For $N_f\mu <h-l$ there exist flat directions along which the
$SO(N_c)$ gauge group is generically broken to $U(1)^{l}$ and
the theory is in a Coulomb phase: There is a collection of $l$ massless
photons in the spectrum.
In these regions, as we shall argue, $R$-symmetry, classical
singularity and the form of the instanton corrections are
sufficient in order to fully determine the quantum moduli spaces.
First recall the curve for $N_f=0$. It reads
\cite{Dan,BL}
\beqar
y^2 &=& {\cal C}^2(x) - \Lambda_0^{2(N_c-2)}x^2 \qquad N_c
\quad odd\comma \nonumber\\
y^2 &=& {\cal C}^2(x) - \Lambda_0^{2(N_c-2)}x^4 \qquad N_c
\quad even\comma
\eeqar
where ${\cal C}(x)$ is given by (\ref{cl}).
\newline
When $N_f$ flavors in the vector representation of
the gauge group are present such that $N_f <h-l$ we have
the following claim:
\newline
{\bf Claim}: The curve describing
the quantum moduli space with
$N_f<N_c-l-2$ flavors is:
\beqar
y^2 &=& {\cal C}^2(x) - \Lambda_{N_f}^{2(N_c-2-N_f)}x^2\det(x^2I_f-M^2)
\qquad N_c\quad odd\comma \nonumber\\
y^2 &=& {\cal C}^2(x) - \Lambda_{N_f}^{2(N_c-2-N_f)}x^4\det(x^2I_f-M^2)
\qquad N_c\quad even\comma
\eeqar
where $I_f$ is the $N_f\times N_f$ identity matrix. Henceforth we will
denote $d=2$ for odd $N_c$
and $d=4$ for even $N_c$. $M$ is the bare
mass matrix of the quarks.
{\bf Proof}:
$R$-symmetry, the form of instanton corrections together with the classical
singularity of the gauge group imply that the most general curve is
\beq
y^2 =  {\cal C}(x)^2-
\Lambda_{N_f}^{2h-2N_f\mu}G(x,m_i) \equiv P \comma
\label{gm1}
\eeq
where $G(x,m_i)$ is a polynomial of degree $2N_f\mu$ in $x$.
A priori, $G$  may also depend on the moduli
$s_k$. In the sequel it will be shown that in fact it is independent
of the moduli.
The second term in (\ref{gm1}) is the quantum correction  to the
classical curve (\ref{cl1}). Note that there are
only one instanton contributions to the curve (\ref{gm1}) due to
the $R$ charge restriction.
The polynomial $G(x,m_i)$ is determined
by requiring that the discriminant
of the polynomial $P$
in (\ref{gm1}) has the right classical
limit (\ref{Del}).
In order to do this we need to evaluate first the classical
limit of the discriminant.
Note that the discriminant of a polynomial can be evaluated up to a
multiplicative constant by
\footnote{The discriminant of a degree $n$
curve is a polynomial in its roots of degree $n(n-1)$ that
vanishes when any two roots of the curve coincide. It is easy to
see that (\ref{discdef}) satisfies these.}
\beq
\Delta[P] = \prod_{x_i \in S}P(x_i) \comma
\label{discdef}
\eeq
where $x_i \in S$ are the critical points of $P$,
$\partial_x P(x_i) = 0$.
Differentiating the polynomial in (\ref{gm1}) with respect
to $x$ and equating to zero we have
\beq
P' = 2{\cal C}(x)
{\cal C}(x)'-
\varepsilon G(x,m_i)' =0 \comma
\label{gm1d}
\eeq
where we denoted $\varepsilon =  \Lambda_{N_f}^{2h-2N_f\mu}$.
The roots of equation (\ref{gm1d}) are $\{r_i=a_i +
\varepsilon~~ corrections \}$ and
$\{s_i = b_i + \varepsilon~~ corrections \}$ where $a_i$ and $b_i$ are
the roots of ${\cal C}(x)$ and ${\cal C}(x)'$, respectively.
According to (\ref{discdef})
\beq
\Delta[P] = \prod_iP(r_i)\prod_jP(s_j) \stop
\label{val}
\eeq
In order to analyze the classical limit
we have to evaluate (\ref{val}) as $\varepsilon \rightarrow 0$.
Consider the first product in (\ref{val}) and let us prove that
to lowest order in $\varepsilon$
\beq
\prod_iP(r_i) = \varepsilon^{h_g}\prod_i G(a_i,m_j)
\comma
\label{val1}
\eeq
which is the contribution of the  $\varepsilon G$ term
in $P$.
We have to show that the  ${\cal C}(x)$ contribution
is of higher order in $\varepsilon$.
Suppose $a_i$ is a root of ${\cal C}(x)$ of multiplicity $n$, then
it follows from  (\ref{gm1d}) that
\beq
r_i = a_i + c_i \varepsilon^{\frac{1}{2n-1}} +\ldots \stop
\label{reps}
\eeq
Thus the contribution of ${\cal C}(r_i)^2$ is of order
$\frac{2n}{2n-1} > 1$ in $\varepsilon$ which is higher than
the contribution of the $\varepsilon G$ term.
In order to get the lowest order in $\varepsilon$
we evaluate the
second product in (\ref{val}) at the roots $b_i$, which yields
using (\ref{discdef})
\beq
\prod_jP(b_j) = \Delta[{\cal C}]^2  \comma
\label{val2}
\eeq
where  $\Delta[{\cal C}]$, given by (\ref{D0}) is the discriminant
of the classical curve (\ref{cl1}). Thus,
(\ref{val1}) and (\ref{val2}) yield
\beq
\Delta[P] = \varepsilon^{h_g}\Delta[{\cal C}]^2
\prod_i G(a_i,m_j) \stop
\label{val3}
\eeq
A comparison of (\ref{val3}) to (\ref{Del}),(\ref{D0})
and (\ref{D1}) yields
\beq
G(x,m_i) =\det(x^2-M^2)\comma
\label{G}
\eeq
which completes the proof.

The meromorphic one-form $\lambda$, satisfying
(\ref{lambda}), takes the form
\beq
\lambda = \frac{x dx}{2\pi iy}\( \frac{{\cal C}
G'}{2G}-{\cal C}' \) \comma
\label{lam}
\eeq
for $2N_f\mu<h$.
The fact that the residues of $\lambda$ satisfy (\ref{res}) is
a corollary of the residue calculation that will be presented
in the next section.
This provides a consistency check on our result.

\section{The quantum moduli space for  $N_f\geq N_c-l-2$}

\subsection{The case $N_f=N_c-l-2$}

The most general curve consistent with $R$-symmetry,
instanton corrections and classical singularity is
\beq
y^2 = {\cal C}^2(x) - \Lambda_{N_f}^{2l}\(x^d\det(
x^2I_f-M^2)+a{\cal C}(x)\)+b\Lambda^{4l},
\label{ncnf}
\eeq
where $a$ and $b$ are constant coefficients.
Since the one instanton correction $\Lambda^{h}$ has $R$ charge
$h$, the curve (\ref{ncnf}) gets contributions also from a two
instanton process of the form $\Lambda^{2h}$.
The structure of the correction  $\Lambda^{h}{\cal C}(x)$
in (\ref{ncnf}) is determined such that it vanishes for $x=a_i$
as required, via the analysis of the previous section,
by comparing
(\ref{val3}) with (\ref{D1}).

The curve (\ref{ncnf}) can be written in a form suitable
for generalization to $N_f>N_c-l-2$:
\beq
y^2 =  F^2 - H \comma
\label{ncnf1}
\eeq
where
\beqar
F & = &  {\cal C} + \alpha \Lambda^{h} \nonumber\\
H & = & \Lambda^{h}\(x^d \det(x^2I_f-M^2) + \beta
\Lambda^{h}\) \comma
\label{ncnf3}
\eeqar
with $\alpha$ and $\beta$ constants.

The meromorphic one-form $\lambda$ (\ref{lambda}) reads in this case
\beq
\lambda = \frac{x dx}{2\pi iy}\( \frac{F H'}{2H}-F' \) \stop
\label{lamn}
\eeq
The residue formula (\ref{res}) can be used in order to determine the
coefficient $\beta$, as we will show now.
This will be an example of a powerful constraint which will be much used
in the sequel.

Let us consider the case of equal bare masses $m_i=m$. The zeros of
$H$ in (\ref{ncnf3}) read
\beq
x_i = -m + e^{\frac{2 \pi i}{N_f}} \beta^{\frac{1}{N_f}}\Lambda
\comma~~~~~~i=1,...,N_f-1 \stop
\label{roots}
\eeq
The residue of $\lambda$ in (\ref{lamn})
at the root $x_0$ of (\ref{roots}) is
\beq
2\pi i\res_{x=x_0}(\lambda) = \frac{m-\beta}{2} \comma
\label{res1}
\eeq
thus (\ref{res}) implies $\beta =0$.
The residues of $\lambda$ at zeros of $y$ vanish.
This is easily seen by differentiating (\ref{ncnf1}) with respect to $x$,
which together with (\ref{ncnf1}) at $y=0$ yield
\beq
\(\frac{F H'}{2H}-F' \)_{y=0} =0 \stop
\eeq
Thus, we verified that the residue formula (\ref{res}) is satisfied
completely by $\lambda$ of (\ref{lamn}).
Note that we are still left with one undetermined constant
coefficient $\alpha$.
This constant may br determined by global symmetry arguments.
We choose to absord this constant by redefining $s_{2l}$.

Compatibility with the hyper-elliptic curves for $N_c>N_f$ which will
be discussed in the next section implies that $\alpha =\frac{1}{4}$
for $N_c>2$, however, we do not have a full proof for that.

Thus, we suggest that the family of curves
describing the quantum moduli space of vacua
for the $N_f=N_c-l-2$ is
\beqar
y^2 &=& {\cal C}^2(x) - \Lambda^{2l} x^d\det(x^2I_f-M^2) \stop
\label{nn}
\eeqar

\subsection{The quantum moduli space for $N_f > N_c-l-2$}

When the number of flavors is increased the curves describing the
quantum moduli spaces may get contributions from higher multi-instanton
processes. This increases the number of terms
in the polynomials describing the curves that should be
determined.
In this section we study the general $N_f>N_c$ cases.
We parameterize the curves and determine there structure up to a
certain number of unknown constant coefficients.
Global symmetry considerations may suggest a fixed value for these
constants. We reabsorb the constants in a redefinition of the symmetric
functions $s_{2k}$.

The general structure of the family of curves describing the
quantum moduli space of vacua when $N_f>N_c-l-2$ is encoded in the
following claim.\\
{\bf Claim}: The curve describing the quantum moduli space with
gauge group $SO(N_c)$ and
$N_f>N_c-l-2$ flavors is:
\beq
y^2 = \({\cal C}(x)+\Lambda^{2(N_c-N_f-2)}P\)^2
-\Lambda^{2(N_c-N_f-2)}x^d\det(x^2I_f-M^2)
\label{nfgnc}
\eeq
where $P(x,m_i,\Lambda)$ is a polynomial of degree $h+2(N_f-N_c+2)$ in
$x,m_i$ and is {\it independent} of the moduli $s_k$.\\
{\bf Proof}: Consider the most general hyper-elliptic curve.
Using $\partial_{2k} \partial_{2l}\lambda=
\partial_{2l} \partial_{2k}\lambda$ where
$\partial_k \equiv \frac{\partial}{\partial s_k}$ together with
(\ref{lambda}) yields
\beq
x^{-2k} \partial_{2l} y^2 =
x^{-2l} \partial_{2k} y^2 \stop
\label{eqh}
\eeq
Equation (\ref{eqh})
implies that $y^2$ depends on the moduli
$s_k$ only via ${\cal C}(x)$.
Thus, $y^2(s_k)= y^2({\cal C}(x)).$
Moreover, $R$ charge symmetry implies that only terms
up to quadratic in  ${\cal C}(x)$ can appear:
\beq
y^2 = {\cal C}^2 g_0 +{\cal C}g_1(x,m_i,\Lambda) +
g_2(x,m_i,\Lambda) \comma
\label{sol}
\eeq
where $g_0$ is a polynomial of degree $0$ in $x$, namely a constant
\footnote{As we will discuss in the sequel,
when $N_f=N_c-2$, $g_0$ may be
a modular form $g_0(q)$.},
$g_1,g_2$ are polynomials in $x$ of degree $h_g$
and $2h_g$, respectively and are
independent of the moduli $s_k$.
The classical limit fixes $g_0=1$. The curve (\ref{sol}) may be
recast in the form
\beq
y^2 = \({\cal C}+\Lambda^{2(N_c-N_f-2)}P\)^2-\Lambda^{2(N_c-N_f-2)}
G \comma
\label{gf}
\eeq
with
\beq
G(x,m_i,\Lambda) = x^d\det(x^2I_f-M^2)+
 \sum_{k=1}^{\[\frac{2N_c}{2(N_c-N_f-2)}\]-1}
\Lambda^{2k(N_c-N_f-2)}n_k(x,m) \comma
\label{nfgnc1}
\eeq
where we used $R$ charge considerations and the form of instanton
corrections.
The first term in (\ref{nfgnc1})
has been deduced from the structure of the
classical singularity in section $3$.
The other terms are arranged according to the
order of multi-instanton contribution.
$n_k$ is a polynomial of degree $(k+1)N_f-2kN_c$ and
$\[\frac{2N_c}{2N_c-N_f}\]$ denotes its integer value.
$P(x,m_i,\Lambda)$ is a polynomial of degree $N_f$ in $x,m_i$.

We will now use the residue formula
(\ref{res}) in order to determine the form of
$G(x)$ in (\ref{gf}).
The meromorphic one-form $\lambda$ (\ref{lambda}) associated with
(\ref{gf}) is
\beq
\lambda = \frac{x dx}{2 \pi i y}\(\frac{FG'}{2G}-F'\) \comma
\label{lam1}
\eeq
where, as before, prime denotes derivative with respect to $x$ and
$F= {\cal C}+\Lambda^{2(N_c-N_f-2)}P$.

Consider the zeros $x_i$ of $G(x)$ where $y(x_i)\neq 0$ and let
us evaluate $\res(\lambda)$ at these points.
Since $F/y =\pm 1$ at $x_i$ and $G'/G$ equal
the order $n$ of the root $x_i$
we get
\beq
2\pi i\res_{x=x_i}(\lambda)= \pm {n\over2}x_i
\stop
\label{gres}
\eeq
Comparing (\ref{gres}) to (\ref{res}) we conclude that the roots of
$G(x)$ do not receive quantum corrections, thus
$n_k=0$ in (\ref{nfgnc1}). This completes the proof of the claim.

In order to fully construct the curve we still have to determine
the polynomial $P$.
Studies of the structure of the singularities,
compatibility with the global
symmetries may determine $P$. We will absorb $P$ into a
redefinition of $G(x)$ which is a linear transformation on $s_{2k}$
which depends on the masses $m_i$ and the dynamical scale $\Lambda$
but does not change their classical limit.
Thus, we suggest that the familiy of curves describing the quantum
moduli space of vacua for $N_f>N_c-l-2$ is
\beq
y^2 = {\cal C}^2(x)
-\Lambda^{2(N_c-N_f-2)}x^d\det(x^2I_f-M^2)
\label{nfgnc2}
\eeq

\section{The quantum moduli space for $N_f=N_c-2$}

\subsection{The general case}

When $N_f=N_c-2$ with the bare masses and all but one of the moduli
set to zero we get conformally invariant theories.
In these cases the classical relations (\ref{uka}) and (\ref{symm})
are expected to be valid quantum mechanically.
Thus,
\beq
a_D^i =\tau_{ij}a^j \comma
\label{tau}
\eeq
where $\tau_{ij}$ is the matrix of theta angles and coupling constants
of the theory.
The classical and quantum moduli spaces are identical and are described
by a hyper-elliptic curve with period matrix $\tau_{ij}$
and
periods as in (\ref{aap}) with  $a^i,a_D^i$ satisfying (\ref{symm}).

The structure of the family of curves describing the moduli space
of vacua for $N_f=N_c-2$ is encoded in the following:
\newline
{\bf Claim}: The curve for the quantum moduli space for $SO(N_c)$
gauge group with
$N_f=N_c-2$ is:
\beq
y^2 = \[{\cal C}(x,l(q)s_k)+P(x,l(q)m,q)\]^2-L(q)
\det(x^2I_f-l(q)M^2) \stop
\label{2nc}
\eeq
$l(q)$ is a modular form satisfying $l(q) \rightarrow 1$ as
$q\rightarrow 0$.
$P(x,l(q)m,q)$ is a modular form satisfying $P \propto q$ as
$q\rightarrow 0$ and a polynomial of degree $N_c$ in
$x$ which is {\it independent} of the moduli $s_k$.
$L(q)$ is a modular form of weight zero satisfying
$L(q) \propto q + O(q^2)$.
\newline
{\bf Proof}: The proof is similar to that of section 5.3.
The main difference is that the dynamically generated scale
$\Lambda$ is replaced by $q$ as defined in (\ref{ql}).
The structure of the first term in (\ref{2nc}) is deduced following the
same argument as in equations (\ref{eqh}) - (\ref{gf}) together with
$R$-symmetry. The factor $g_0$ of (\ref{sol}) gets contributions both
from ${\cal C}(x,l(q)s_k)$ and $P(x,l(q)m,q)$.
The modification of all the moduli $s_k$ by the same modular
form $l(q)$ is consistent with $R$-symmetry as well as (\ref{eqh})
and (\ref{sol}).

The structure of the second term follows from the analysis
of the residues (\ref{gres}) of the meromorphic one-form $\lambda$
(\ref{lam1}), which now gets a pre-factor $\frac{1}{l(q)}$.
This implies that $m_i$ must be rescaled by $l(q)$ in order
to ensure that the residues of $\lambda$ be independent of $\tau$.
The behavior of $L(q)$ as $q\rightarrow 0$
is implied by the matching condition (\ref{ql})
when integrating a massive quark. It is computed by requiring the curve
to have the classical period matrix.

$P$ may determined by global symmetries and singularity structure.
We choose to absorb it in a redefinition of ${\cal C}(x)$.
This amounts to shifts in $s_{2k}$ by functions which depend on $q$ and
$m_i$ but do not affect their classical limit.
Thus we propose the curve
\beq
y^2 = {\cal C}^2(x,l(q)s_k)-L(q)x^d
\det(x^2I_f-l(q)M^2) \stop
\label{suggest}
\eeq

Consider now the curve (\ref{suggest})
with the masses being set to zero
and let us use the complex line $l$ defined by
$s_i=0, i\neq h,s_{h}=s$.
This curve contains only one scale therefore it is scale invariant.
$L(q)$ will be determined by requiring the curve to have the classical
period matrix.

\section{Discussion and conclusions}

In this paper we
constructed the hyper-elliptic curves which describe the quantum
moduli spaces of vacua of $N=2$ supersymmetric $SO(N_c)$ gauge theories with
$N_f$ flavors of quarks in the vector representation.
We showed that
the curves for $N_f<N_c-l-2$ are completely determined by $R$-symmetry, the
form
of instanton corrections and the requirement
for the correct classical singularity structure.
The compatibility of the residue calculations with the
BPS formula as well as the correct weak coupling monodromy
provide further support to the results.

Weak coupling monodromies are computed along the same lines of
reference \cite{HO} for all the
curves and can be shown to coincide with what is expected on physical
grounds, thus providing a check on the results.
We left the calculation of the strong coupling monodromies
for the future. This will clearly be needed in order
to extract the physics of these theories.
Along the way we derived the
exact metrics on the quantum moduli spaces as well as the exact
spectrum of stable massive states.

The ability to extract exact results in the theories that were studied
in this paper points to an underlying integrable
structure \cite{Mor}. In particular, one expects that the pre-potential
will be
related to
a $\tau$ function of some integrable hierarchy and that the variety
describing the quantum moduli space of vacua
will arise as a solution to a non-linear
integrable equation of the hierarchy. Revealing these structures
may provide us with powerful computational tools
for these four-dimensional models. See for example \cite{NT}.
An Example of a strong consistency check would be the calculation of
the Witten index from the curves and then compare it to the expectation
from weak coupling behavior for large masses of the adjoint and the
quark fields.

In a recent paper Martinec and Warner showed that the Riemann surfaces
which describe the moduli space of quantum vacua for $N=2$
supersymmetric Yang-Mills theories with a simple gauge group $G$ is
described by spectral curves of the periodic Toda lattice for the
dual group. Their solution coincides with the known curves so far and
provides new solutions for the unknown cases. The curves read
\beqar
&A_l& : z+{\mu\over z}+{\cal C}(x)=0,	\nonumber\\
&B_l& : x(z+{\mu\over z})+{\cal C}(x)=0,	\nonumber\\
&C_l& : (z-{\mu\over z})^2+{\cal C}(x)=0,	\nonumber\\
&D_l& : x^2(z+{\mu\over z})+{\cal C}(x)=0
\label{MWcur}
\eeqar
and similar expressions for the exceptional groups.
There exist simple tranformations between the forms of $A_l, B_l, D_l$
and the known curves in the literature \cite{MW}.
There is no known expression for the $C_l$ groups in terms of
hyper-elliptic curves.

The parameter $\mu$ for $N_f=0$ in equation (\ref{MWcur}) is
$\mu=\Lambda^{2h}$. $h$ being the second Casimir of the group.
Henceforth we will denote the Dynkin index of a representation by
$T(r)$.
The previous works on $A_l$ theories with $N_f$ flavors in the
fundumental representation \cite{HO,APS} give
$\mu=\Lambda^{2h-2N_fT(r)}\det(xI_f+M)$ for any flavor.
(The masses are rescaled by a factor which depend on $\tau$ for the
finite theories.)
This work suggests that for $B_l$ and $D_l$ groups with $N_f$ flavors
in the vector representation
$\mu=\Lambda^{2h-2N_fT(r)}\det(x^2I_f-M^2)$.
On the basis of this we conjecture that to any $N_f$ for $C_l$ groups
\beq
\mu=\Lambda^{2h-2N_fT(r)}\det(x^2I_f+M^2).
\eeq
Similar formulae are argued for the exceptional groups.
$\mu=\Lambda^{2h-2N_fT(r)}\det(x^\alpha-M^\alpha),$ Where $\alpha$ is
an integer which depends on the group and on the representation.

\section*{Acknowledgements}
I would like to thank K.~Intriligator and E.~Witten for helpful
discussions.

\end{document}